\documentclass[preprint,12pt]{elsarticle}

\usepackage{graphicx}
\usepackage{color}
\usepackage{url}
\usepackage{multirow}
\usepackage{enumerate}
\usepackage{graphicx}
\usepackage{braket}
\usepackage{hyperref}
\usepackage{amsmath}
\usepackage{physics}
\usepackage{subfig}

\usepackage{mathtools}
\usepackage{multirow}
\usepackage{mathrsfs}
\usepackage{amsfonts}
\usepackage{mathtools}
\usepackage{comment}
\usepackage{longtable}
\usepackage{multicol}
\usepackage{wrapfig}
\usepackage{lscape}
\usepackage{rotating}
\usepackage{booktabs}

\DeclareUnicodeCharacter{2212}{-}

\usepackage{mathtools}

\usepackage{algorithm,algpseudocode}
\algrenewcommand\algorithmicrequire{\textbf{Input:}}
\algrenewcommand\algorithmicensure{\textbf{Output:}}

\begin{document}

\begin{frontmatter}



\title{FragQC: An Efficient Quantum Error Reduction Technique using Quantum Circuit Fragmentation}


\author[1,2]{Saikat Basu\corref{cor1}}
\ead{saikat.basu.000@gmail.com}

\author[1,3]{Arnav Das}
\ead{arnav.das88@gmail.com}

\author[1,4]{Amit Saha}
\ead{abamitsaha@gmail.com}

\author[1]{Amlan Chakrabarti}
\ead{acakcs@caluniv.ac.in}

\author[5]{Susmita Sur-Kolay}
\ead{ssk@isical.ac.in}

\affiliation[1]{organization={A.K.Choudhury School of Information Technology, U. of Calcutta}, city={Kolkata}, country={India}}

\affiliation[2]{organization={LTIMindtree Ltd.}, city={Kolkata}, country={India}}

\affiliation[3]{organization={Dept. of Computer Science, St. Xavier's University}, addressline={}, city={Kolkata}, country={India}}

\affiliation[4]{organization={Eviden (an Atos business)}, city={Pune}, country={India}}

\affiliation[5]{organization={Advanced Computing and Microelectronics Unit, Indian Statistical Institute}, city={Kolkata}, country={India}}

\cortext[cor1]{Corresponding author}




\begin{abstract}
Quantum computers must meet extremely stringent qualitative and quantitative requirements on their qubits in order to solve real-life problems. Quantum circuit fragmentation
techniques divide a large quantum circuit into a number of sub-circuits that can be executed on the smaller noisy quantum hardware available. However, the process of quantum circuit fragmentation involves finding an ideal cut that has exponential time complexity, and also classical post-processing required to reconstruct the output. In this paper, we represent a quantum circuit using a  weighted graph and propose a novel classical graph partitioning algorithm for selecting an efficient fragmentation that reduces the entanglement between the sub-circuits along with balancing the estimated error in each sub-circuit. We also demonstrate a comparative study over different classical and quantum approaches of graph partitioning for finding such a  cut. We present {\it FragQC}, a software tool 
 that cuts a quantum circuit into sub-circuits when its error probability exceeds a certain threshold. With this proposed approach, we achieve an increase of fidelity by 14.83\%  compared to direct execution without cutting the circuit, and 8.45\%  over the state-of-the-art ILP-based method, for the benchmark circuits. 
\end{abstract}

\begin{keyword}
Quantum circuit fragmentation \sep Hybrid quantum systems \sep Quantum error \sep  Graph partitioning \sep Genetic algorithm\sep Circuit cutting \sep Quantum annealing.\\

 {\small The code for {\it FragQC} is available at \href{https://github.com/arnavdas88/FragQC}{https://github.com/arnavdas88/FragQC}.}
\end{keyword}

\end{frontmatter}

\section{Introduction}\label{sec1}

Modern researchers have been developing quantum algorithms to achieve asymptotic improvement with the advance of the technology for quantum computing over the past two decades \cite{chuang}.  
Through qubits and quantum gates, a  quantum algorithm may be implemented as a quantum circuit. Despite the enormous potential of quantum algorithms, the current generation of quantum computers is more error-prone, which restricts their capacity to solve computational problems \cite{Preskill2018quantumcomputingin}.

The use of quantum error correcting codes (QECCs) \cite{10.1007/3-540-46796-3_23, 10.1007/3-540-49208-9_27, Baek_2019,Leuenberger_2001, PhysRevA.52.R2493, PhysRevLett.77.793, Wootton_2020, RevModPhys.87.307, PhysRevLett.77.198} to eliminate noise-related errors opens the door to fault-tolerant quantum computation \cite{Campbell_2017}. However, in reality, implementing quantum error correction entails a significant cost due to the high number of qubit requirements, which is still outside the scope of current near-term devices. Numerous error reduction approaches have been developed as a result of the typical error rate of present near-term technologies; one of them is quantum error mitigation (QEM) \cite{Nautrup_2019,9226505}. QEM does not use additional quantum resources; instead, it uses a variety of techniques, including extrapolation, probabilistic error cancellation, quantum subspace expansion, symmetry verification, machine learning, etc., to improve the accuracy of estimating the outcome in a particular quantum computational problem slightly. According to the most recent research, QEM is constrained to quantum circuits with a small number of qubits and a small depth because of the significant overhead of classical computational time complexity \cite{Nautrup_2019}.

Fragmentation of quantum circuits \cite{Bravyi_2016, PhysRevLett.125.150504, tang2022scaleqc} can be a helpful strategy for overcoming the technical difficulties of QEM since it partitions a quantum circuit into smaller sub-circuits, with fewer qubits and smaller depth. The short coherence periods of noisy intermediate-scale quantum (NISQ) processors must thus be handled by the sub-circuits. When running on a NISQ device, each sub-circuit experiences less noise. On a small quantum computer, a larger quantum system is primarily simulated through the fragmentation of a quantum circuit. In \cite{10.1145/3445814.3446758, 9155024}, the authors suggested fragmenting a quantum circuit to reduce the exponential post-processing cost. Researchers also investigated for the first time in another work \cite{ayral2021quantum} how such fragmentation impacts the various quantum noise models.  Circuit fragmentation was later examined in another work as a way to lessen the impacts of noise \cite{Perlin2021Quantum}. Although the main goal of the earlier research was to break up complex circuits so that they could be implemented, the issue of noise that might cause false results was never appropriately addressed.

\textbf{Motivation:} While splitting a quantum circuit into smaller sub-circuits can lessen the impact of noise, fragmentation drives up post-processing costs, and overall computational costs. In {\it i-QER} \cite{10.1145/3539613}, by predicting error in a quantum circuit while keeping in mind that the fragmentation of a particular quantum circuit is minimized, error reduction of a specific circuit is accomplished. It is not simple to predict error in a quantum circuit. However, in {\it i-QER}, a machine learning-based strategy has been employed to train the system to accurately predict errors in quantum circuits by taking into account their characteristics. Albeit there are several problems with {\it i-QER}, which are (i) the machine learning-based approach has a scalability issue when the circuit size is large, especially with training accurately, (ii) it is always a hardware-dependent method while using a machine learning approach, (iii) the choice of an appropriate machine learning models from a number of those is still an open question.

The aim of this paper is to address all the above-mentioned issues by proposing a generalized circuit fragmentation approach by optimizing both the success probability and the cut size in a graph-theoretic manner without machine learning.

{\bf Our major contributions} are:
\begin{itemize}
    \item Error influenced balanced circuit bi-partitioning algorithm is proposed which not only balances the error in the two partitions but also reduces the inter-partition communications.
    \item We provide a tool named {\it FragQC} to efficiently fragment a quantum circuit with different approaches for achieving higher fidelity of the output quantum state.
\end{itemize}

The structure of this paper is as follows. Section \ref{sec2} briefly presents the preliminary concepts of quantum circuit fragmentation, graph partitioning algorithms, and available quantum hardware. Section \ref{sec3} describes the tool {\it FragQC}. Section \ref{sec4} briefly discusses the experimental results of the proposed methodology. Section \ref{sec5} captures our conclusions.

\section{Background}\label{sec2}

This section briefly explains quantum circuit fragmentation and its challenges followed by a few relevant existing heuristic and approximate approaches for solving graph partitioning problems. Lastly, we shed some light on existing quantum hardware and their error rates.

\subsection{Quantum circuit fragmentation}\label{sec2.1}
A theoretical overview of quantum circuit fragmentation and an illustrative example are presented.  

\subsubsection{The idea}
Each line for a qubit of a quantum circuit represents a  sequence of single and multi-qubit gate operations. The time flows from left to right in the quantum circuit diagram. Quantum Circuit fragmentation cuts these notional qubit wires vertically. 

In \cite{PhysRevLett.125.150504}, the authors demonstrated mathematically that the idea to cut a qubit wire is based on the notion that, if we have multiple copies of an experimentally generated single qubit state with a density matrix $\rho$, then the set $ \{I/\sqrt{2}, X/\sqrt{2}, Y /\sqrt{2}, Z/\sqrt{2}\}$ forms an orthonormal set of matrices with respect to the Hilbert–Schmidt inner product. So $\rho$ may be expanded as:

\begin{equation}\label{frag1}
    \mathbf{\rho}=\frac{Tr(\mathbf{\rho})I+Tr(\mathbf{\rho}X)X+Tr(\mathbf{\rho}Y)Y+Tr(\mathbf{\rho}Z)Z}{2}.
\end{equation}

In order to run on quantum computers, the Pauli matrices can be further decomposed into their eigenbases \cite{10.1145/3445814.3446758} as follows:

\begin{equation}
    \mathbf{\rho}=\frac{\rho_1+\rho_2+\rho_3+\rho_4}{2}
\end{equation}
where
\begin{eqnarray*}
    \rho_1&=&[Tr(\mathbf{\rho}I)+Tr(\mathbf{\rho}Z)]\ket{0}\bra{0}\\
    \rho_2&=&[Tr(\mathbf{\rho}I)-Tr(\mathbf{\rho}Z)]\ket{1}\bra{1}\\
    \rho_3&=&Tr(\mathbf{\rho}X)[2\ket{+}\bra{+}-\ket{0}\bra{0}-\ket{1}\bra{1}]\\
    \rho_4&=&Tr(\mathbf{\rho}Y)[2\ket{+i}\bra{+i}-\ket{0}\bra{0}-\ket{1}\bra{1}]
\end{eqnarray*}

Physically, the trace operators are equivalent to measuring the qubits in one of the  Pauli bases $\sigma_i \in \{I, X, Y, Z\}$, and the density matrices correspond to physically initializing the qubits to one of the eigenstates.

If we assume that a qubit wire connecting $A$ and $B$, two vertices representing gates,  is cut then we can reconstruct the sub-circuits by summing over the four pairs of measurement circuits added to $A$ and an initialization circuit added to $B$. After that, the overall circuit output is reconstructed by adding the four pairs of Kronecker products between the sub-circuit outputs. As a result, there are $4^{k}$ Kronecker products to be computed for a cut size of $k$.

\begin{figure}[!ht]
\centering
\includegraphics[width=\textwidth]{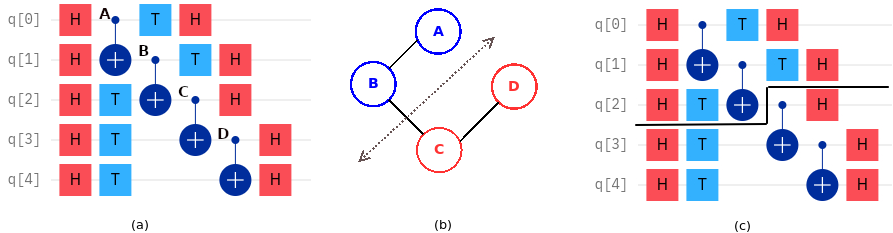}
\caption{Example of quantum circuit fragmentation: (a) a  quantum circuit $C$ with 5 qubits and 4 two-qubit gates; (b) the corresponding graph $G$ of $C$; (c) the two quantum sub-circuits after a fragmentation.}
\label{frag}
\end{figure}

\paragraph{Example of quantum circuit fragmentation}
  Let us consider a quantum circuit with five qubits and four two-qubit {\it CNOT} gates shown in Figure \ref{frag}(a). All the qubits are initialized to $\ket{0}$. In order to implement quantum circuit fragmentation, first we need to construct a graph from the given circuit, where the vertices are the two-qubit gates and there is an edge if two to-qubit gates have at least one in common.  Thus, for a given quantum circuit $C$, we have a graph $G(V,E)$. The task is to find a cut that can separate the vertices into more than one disjoint set as shown in Figure \ref{frag}(b).

The circuit $C$ can be partitioned into two sub-circuits, as shown in Figure  \ref{frag}(b). Here, the two partitions $\{A,B\}$ \& $\{C,D\}$ are separated by the dashed arrow line. The number of edges between the two partitions can be defined as the cut size ($k$). Considering that only one qubit, i.e., the third qubit (q[2]) is connecting the partitions, the value of $k$ is 1. Therefore, each sub-circuit can be executed on a $3$-qubit quantum hardware instead of a $5$-qubit  one. We need to take measurements on the 3$^{rd}$-qubit after the {\it CNOT} (node $B$). The initialization of the sub-circuit $\{C,D\}$ has to be done based on the measurements after $B$. Therefore, conventionally in the classical post-processing step, the complete probability distribution for the entire circuit can be reconstructed by taking the corresponding outputs of the two smaller sub-circuits, running four pairs of Kronecker products, and adding them together.

In \cite{10.1145/3445814.3446758, tang2022scaleqc}, the authors have demonstrated efficient ways for the classical reconstruction method. In \cite{Perlin2021}, the authors have proposed maximum-likelihood fragment tomography (MLFT) as an improved circuit fragmentation technique, with a limited number of qubits to run the quantum sub-circuits on quantum hardware. MLFT further finds the most likely probability distribution for the output of a quantum circuit, with the measurement data obtained from the circuit’s fragments, along with minimizing the classical computing overhead of circuit fragmentation methods. Hence, they showed that circuit fragmentation as a standard tool can be used for running the sub-circuits on quantum devices by estimating the outcome of a partitioned circuit with higher fidelity as compared to the full circuit execution.

\subsubsection{Challenges of quantum circuit fragmentation}

In spite of this immense potential, quantum circuit fragmentation faces a few formidable challenges when it is applied to large quantum circuits. Finding an efficient cut location is the first difficult task. Quantum circuits can always be divided into smaller sub-circuits, but choosing an efficient cut is critical for reducing the amount of classical post-processing and the effects of noise. Partitioning a large quantum circuit into sub-circuits often requires multiple edges or qubit wires to be cut. In such cases, all the possible measurement and initialization combinations have to be evaluated. Hence, the number of Kronecker products required is $4^{k}$, with $k$ being the number of edges cut. Thus, quantum circuits with $n$ edges have a combinatorially explosive search space of $\mathcal{O}(n!)$ to find an efficient cut. 

Additionally, if we consider the effects of noise and look to improve the fidelity of a quantum circuit using the quantum circuit fragmentation approach, the problem of finding an efficient cut becomes even more complex. Section \ref{secthree} addresses this problem with different classical as well as quantum annealing-based approaches. Before that, let us describe a few popular graph partitioning algorithms very briefly.

\subsection{Graph partitioning algorithms }\label{sec2.2}


In this paper, we consider a balanced bi-partition problem for quantum circuit fragmentation. In balanced bi-partitioning a graph, the goal is to partition the graph into two subgraphs with nearly equal disjoint sets of vertices, while minimizing the capacity of the edges between the two subgraphs.
For the sake of completeness, we first describe the popular heuristic algorithms for graph partitioning such as Kernighan-Lin (KL) algorithm \cite{6771089} and Fiduccia–Mattheyses (FM) algorithm \cite{1585498} in brief.  We also discuss the {\it h-METIS}, one of the most popular hypergraph partitioning methods. Further, we also describe a genetic algorithm and a quantum annealing-based method.

\paragraph{Kernighan-Lin algorithm}\label{KL} KL algorithm is a greedy heuristic that tries to identify the optimal bi-partition of a graph. Given a graph, it starts with an initial bi-section, and swaps an equal number of vertices between the two partitions, aiming to improve the cut size over the initial partition. This process is iterated in search of an optimal partition.


While it is a widely used graph partitioning algorithm, it also has some limitations
\begin{itemize}
    \item For large graphs with multiple optimal partitions, this algorithm tends to converge to any of the optimal solutions, depending on the initial state. Hence, instead of converging into the global optima, it may get stuck in the local optima.
    \item The quality of the initial partition significantly influences the final partition obtained by the KL algorithm. A poorly chosen initial partition may lead to sub-optimal results or longer convergence time. Finding a suitable initial partition can be difficult for large graphs.
    \item Each pass of the KL algorithm takes $\mathcal{O}(n^3)$ time, which makes it computationally expensive for large graphs. Thus it lacks scalability as well.
\end{itemize}

\paragraph{Fiduccia–Mattheyse(FM) algorithm} FM algorithm is an improved version of the KL algorithm which tries to overcome the limitations of the KL algorithm. Unlike the KL algorithm, the FM algorithm does not swap two vertices among the partitions. It calculates the gain of each vertex in its initial partition and moves the vertex with the highest gain to the other partition. It is the wizardry of a doubly linked list implementation of the algorithm, which makes it a linear time algorithm. However, in order to achieve a lower cut size, the FM algorithm allows imbalanced partitions to a certain degree. Both algorithms do not explicitly enforce any constraint to ensure balanced partitioning. 

\paragraph{h-METIS} {\it h-METIS}  proposed by G. Karypis et al. in \cite{doi:10.1137/S1064827595287997,KARYPIS199896}that partitions large hypergraphs, mostly generated while circuit design. The concept of {\it h-METIS}  is based on multilevel graph partitioning, as explained in \cite{10.1145/309847.309954,10.1145/266021.266273}. Unlike other graph partitioning algorithms, {\it h-METIS}  does not perform partitioning operations on the original graph. It takes a coarsening approach repeatedly, where the vertices and edges of the given graph are collapsed to reduce it to a smaller graph. It then performs the partitioning operation on the small graph. In the next phase, it performs an uncoarsening on the two subgraphs obtained along with refinements to the partitioning. In this manner, {\it h-METIS}  can quickly produce high-quality partitions for a large variety of hypergraphs. Experiments performed in \cite{KARYPIS199896} on a large number of hypergraphs show that {\it h-METIS}  produces consistently better partitions than those by other widely used algorithms, such as KL, FM, etc. We leverage the consistent tool for circuit partitioning to compare with our approach in our multiple experiments.

\paragraph{Genetic algorithm} Genetic algorithm (GA) is a metaheuristic inspired by natural selection. GAs, which rely on biologically inspired operators such as selection, crossover, and mutation, are often employed to develop near-optimal solutions to optimization and search problems. A GA initializes with an initial set of solutions or an initial population, which evolves into different populations with each iteration or generation. After multiple generations, the algorithm returns the best member of the population as the solution to the given problem. 

After each generation, two members of the population are chosen and are then combined to create offspring by using a crossover operator. Mutation operator further modifies an offspring with a very low probability, to include more diversity in the population.

In \cite{508322}, the authors have applied a GA for the graph bi-partitioning problem. They also claim that the GA performs comparably to or better than KL, FM, and simulated annealing algorithms. Thus, we plan to use the basic idea of the GA to construct our classical approach for solving graph partitioning problem.


\paragraph{Quantum annealing}\label{2.3}
The Hamiltonian of a system represents the total energy of the system. If the Hamiltonian of a system is very slowly evolved from an initial state to a final state, then the adiabatic theorem \cite{Born1928} states that, if the system is in the $n^{th}$ eigenstate of the initial Hamiltonian, it evolves as the $n^{th}$ eigenstate of the final Hamiltonian.  

In \cite{farhi2000quantum}, the authors proposed a quantum annealing algorithm that leverages this adiabatic evolution theorem to solve different combinatorial optimization problems. In the quantum annealing (QA) algorithm, we encode the solution to the problem in the ground state of a Hamiltonian. Therefore, we choose the initial state of the system to be the ground state of a simple Hamiltonian. This initial Hamiltonian is then slowly evolved to the final Hamiltonian whose ground state encodes the solution to the problem.
Therefore, if the evolution is slow enough, then according to the adiabatic theorem, the system remains in its ground state throughout the evolution, and we have the solution to our problem in the final state.

If we can encode the objective function of a graph partitioning problem into a Hamiltonian, then the QA can find the optimal partitioning strategy. In this paper, the graph partitioning algorithm is used for balancing the error of a quantum circuit into two quantum sub-circuits. Thus before taking a look into the proposed methodology, it is important to discuss the quantum hardware and their error rates.



\subsection{Quantum hardware and its error rate}\label{sec2.4}

\begin{figure}[!ht]
\centering
\includegraphics[scale=0.4]{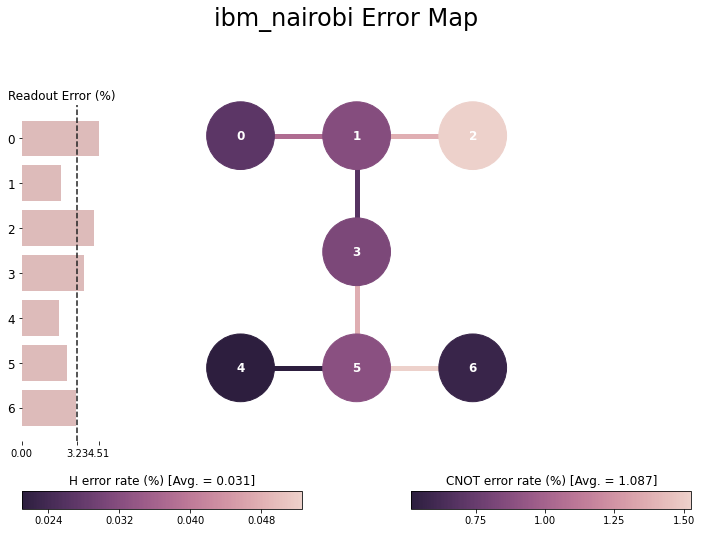}
\caption{Error Map of 7-qubit $ibm\_nairobi$ device.}
\label{error_map}
\end{figure}

Superconducting quantum devices \cite{PhysRevA.76.042319}, quantum dots \cite{QD}, ion traps \cite{Bruzewicz_2019}, and neutral atoms \cite{NA} are currently the most popular quantum technologies for constructing qubits and quantum gates. For hardware compatibility, the quantum logic gates in a quantum circuit, such as {\it CNOT, Hadamard, S, T, X, Y, Z} must be decomposed using primitive gate operations of a specific quantum hardware or NISQ (Noisy Inter-mediate Scale quantum) \cite{Preskill2018quantumcomputingin} device. Further, each quantum device has a dedicated qubit connectivity topology as each is built with its own unique set of physical qubits, coupling strengths, and control mechanisms. These variations can result in different noise characteristics. Some devices, for example, may contain qubits that are closer to one another, resulting in stronger interactions and perhaps greater crosstalk between qubits. Thus quantum hardware has hardware-specific properties such as gate error rates, readout errors, etc. In this paper, we consider IBM's superconducting-based hardware for further experiments. Figure \ref{error_map} depict the error map of $ibm\_nairobi$ . It shows the qubit connectivity layout as well as the error rates for different gate operations in the qubits.

 
Along with error rates of different gate operations, there are a few critical hardware-specific properties such as relaxation time, coherence time, etc. The relaxation time $T_1$ is a crucial parameter which is a measure of how long a qubit can remain in a superposition state or maintain its quantum information before decohering into a classical state. The coherence time $T_2$ represents the duration during which a quantum system, can preserve the quantum phase information, that enables quantum computations.

Therefore, there are multiple quantum devices available to perform our experiments, although these can be performed on any other available quantum hardware. We perform our experiments on a quantum device with a sufficient number of qubits so that we can easily execute medium to large quantum circuits. Thus we choose $ibm\_Sherbrooke$, which is a 127 qubit quantum system with median $T_{1}$ and $T_{2}$ of 295.33 $\mu$s and 166.02 $\mu$s respectively. 




\section{Our Proposed Tool: \textit{FragQC}}\label{sec3}
\begin{figure}[!ht]
\centering
\includegraphics[scale=0.45]{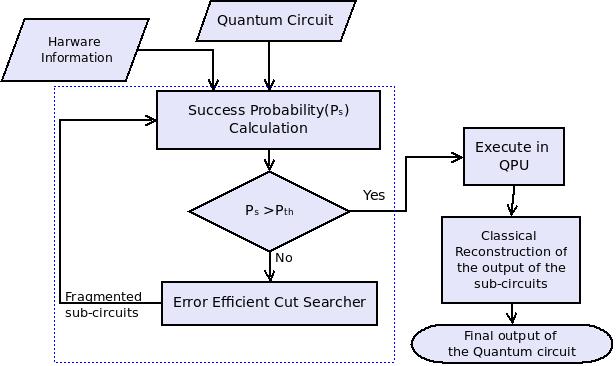}
\caption{Flowchart of the proposed tool {\it FragQC}.}
\label{overall_flow}
\end{figure}

An overview of {\it FragQC}, our technique for reducing quantum errors, is provided below. A flowchart of the suggested tool containing the essential modules is given as Figure \ref{overall_flow}. 

The tool accepts a quantum circuit as an input and calculates its potential success probability while considering the noise profile of a specific hardware. A novel error influenced balanced bi-partitioning algorithm is executed to partition the circuit into two sub-circuits whenever the projected success probability is below a certain user-specified threshold.
This bi-partitioning is continued recursively until the sub-circuits can be run with a reasonable chance of success. For the sake of simplicity, we consider the success probability threshold of each sub-circuit akin to the success probability threshold of the overall circuit, which is given by the user. After the sub-circuits are implemented on the hardware, their outputs (probability distributions) are combined appropriately to generate the output of the entire circuit. We have adopted the output reconstruction method from \cite{10.1145/3445814.3446758}.

\subsection{Proposed technique on circuit fragmentation}\label{secthree}

\begin{figure}[!ht]
\centering
\includegraphics[scale=0.45]{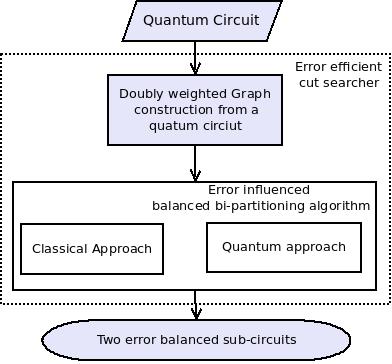}
\caption{Block diagram of our error efficient cut searcher.}
\label{BD_FragQC}
\end{figure}

The proposed method of circuit fragmentation, shown in Figure \ref{BD_FragQC}, has two main parts, namely (i) constructing a doubly weighted graph for a given quantum circuit, and (ii) error influenced balanced partitioning algorithm for circuit fragmentation.

\subsubsection{Graph representation of a quantum circuit}\label{sec3.1}
First, we represent the given quantum circuit $C$ as a graph $G_C$. We  denote each two-qubit gate in $C$ as a vertex, and an edge between two vertices indicates that the corresponding two-qubit gates share one or two qubits. The weight of the edge is either 1 or 2 depending on the number of qubits shared by the two two-qubit gates.  

Our goal is to cut the quantum circuit in such a way that it not only decreases the interaction between the two partitions but also balances the probability of error in each partition. When executed on quantum hardware separately, this reduces the impact of noise. In order to ensure this, we intend to store the error probability information of the circuit for specific quantum hardware as the vertex weight of the graph. Before the details of the calculation of the vertex weight from the error probabilities are given, we briefly discuss the quantum error model.

\paragraph{Quantum error model} Quantum errors can be broadly categorized into two major types: errors due to noisy gate operation and errors due to idle qubits. The noise model can be expressed in terms of the Kraus operators \cite{10.5555/1972505}. Let us consider a pure state $\psi$, and its density matrix $\sigma =\ket{\psi}\bra{\psi}$. The evolution of the state $\psi$ in a quantum channel can be given by a function $\xi$ of its density matrix $\sigma$given as 
\begin{equation}\label{Kraus}
    \xi(\sigma)=\sum_{i} K_{i}\sigma K^{\dag}_{i},     
\end{equation}
where $K_i$ is the Kraus operator and $K^{\dag}_{i}$ is complex conjugate transpose of $K_i$. The evolution of a noisy quantum system can also be represented by Eqn. \ref{Kraus}. If we consider depolarizing noise channels, then the Kraus operators are the Pauli matrices. 

\textit{1. Gate operation error:} The possible error model for a quantum system with one qubit can be expressed as:

\begin{equation}
    \xi(\sigma)=\sum_{i\in \{0,1\}} \sum_{j\in \{0,1\}} p_{i,j} (X^{i}Z^{j})\sigma (X^{i}Z^{j})^{\dag}.
\end{equation}
where $X$ and $Z$ are Pauli operators, and the probability of the corresponding Kraus operator is denoted by $p_{i,j}$. 

Hence, the possible quantum error channels are 
\begin{itemize}
    \item[] if $i=0$ and $j=0$ then $X^{0} Z^{0}=I$, i.e., no error
    \item[] if $i=0$ and $j=1$ then $X^{0} Z^{1}=Z$, i.e., phase flip error
    \item[] if $i=1$ and $j=0$ then $X^{1} Z^{0}=X$, i.e., bit flip error
    \item[] if $i=1$ and $j=1$ then $X^{1} Z^{1}=XZ$, i.e., both bit and phase flip errors
\end{itemize}
In \cite{PhysRevA.86.032324}, the authors represented a noisy gate operation by the ideal gate operation followed by a set of Pauli operators $\{X,Y,Z\}$ with probability $p_{ex}$, $p_{ey}$ and $p_{ez}$ respectively. 

\textit{2. Amplitude damping error:} When a qubit in an open quantum system is kept idle, it can absorb or dissipate energy and change its state spontaneously over time. Let us assume that a qubit can dissipate and absorb energy with a probability $p$ and $(1-p)$ respectively. This noise channel is called an amplitude damping channel and it can also be defined using Kraus operators. The Kraus operators for energy dissipation or state change $\ket{1} \rightarrow \ket{0}$ are written as: 

\[
K_0=\sqrt{p}
\begin{bmatrix}
    1  &  0      \\
    0  &  \sqrt{1-\lambda}      
\end{bmatrix}
,\hspace{1.5mm} K_1= \sqrt{p}
\begin{bmatrix}
    0  &  0      \\
    0  &  \sqrt{\lambda}      
\end{bmatrix} 
,
\]
The Kraus operators for energy absorption or state change $\ket{0} \rightarrow \ket{1}$ are written as: 

\[
K_2=\sqrt{1-p}
\begin{bmatrix}
    \sqrt{1-\lambda}  &  0      \\
     0 &  1      
\end{bmatrix} 
\& \hspace{1.5mm}K_3=\sqrt{1-p}
\begin{bmatrix}
    0  &  0      \\
    \sqrt{\lambda}  &  0      
\end{bmatrix}.
\]
Here, $\lambda \propto e^{-\tau/T_1}$, where $T_1$ is called the energy relaxation time of the quantum system and  $\tau$ is the time duration of the quantum system or the time duration for which the quantum circuit is operational. 

\textit{3. Phase damping error:} Phase damping is a unique quantum mechanical noise model that describes the loss of quantum coherence without loss of energy. The Kraus operators for the dephasing channel can be given as 
\[
K_{p0}=\sqrt{p}
\begin{bmatrix}
    1  &  0      \\
    0  &  \sqrt{1-\lambda}      
\end{bmatrix}
,\hspace{1.5mm} K_{p1}= \sqrt{p}
\begin{bmatrix}
    0  &  0      \\
    0  &  \sqrt{\lambda}      
\end{bmatrix} .
\]
Here, $\lambda \propto e^{-\tau/T_2}$, where $T_2$ is called coherence time of the quantum system and $\tau$ denotes the time duration.

\paragraph{Computing the vertex weights} 
In \cite{majumdar2016error}, the authors have given a linear time algorithm to trace error in a quantum circuit, and in \cite{majumdar2021optimizing,PhysRevA.105.062453}, the authors have computed the probability of success for a quantum circuit considering errors due to noisy gate operations and amplitude damping. Inspired by these works, we first compute the error probability of a quantum circuit $C$. We consider the most common error model for quantum systems, namely   errors due to gate error and idle error \cite{10.1145/3307650.3322253} modeled with amplitude and phase damping error. Let us assume that $C$ has $k_1$ single-qubit gates,  $k_2$  two-qubit gates, and the error probability of a single-qubit gate and a two-qubit gate is $p_1$ and $p_2$ respectively. The error probability of the circuit due to the noisy gate operations can be written as
\begin{equation}
    p_{GE}=1-\{(1-p_1)^{k_1}(1-p_2)^{k_2}\}.
\end{equation}
The error due to the amplitude damping and phase damping along with the gate error can therefore be expressed as

\begin{equation}\label{error}
    P_{Error}=1-\{(1-p_1)^{k_1}(1-p_2)^{k_2} exp^{-(\tau/T_1+\tau/T_2)}  \}.
\end{equation}

The weight of a vertex from   Eqn. \ref{error} is the error of the corresponding two-qubit gate, and the product of the errors of the sequence of single-qubit gates operating on the qubit responsible for the edge. We also consider the idle error for those edges, i.e., decoherence prior to that two-qubit gate. We normalize the calculated error probability to improve accuracy and assign it to the vertex as weight.  

\begin{figure}[!htb]
    \centering
    \subfloat[\centering A  quantum circuit  $C$.]{{\includegraphics[width=\textwidth]{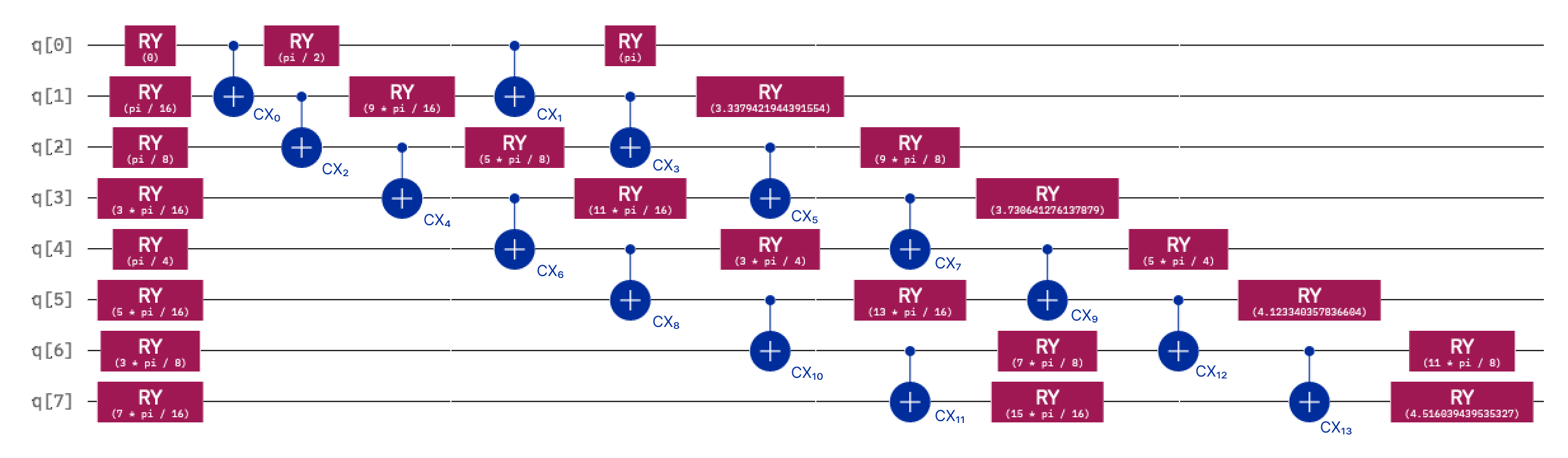} }}%
    \qquad
    \subfloat[\centering The doubly-weighted graph $G_C$ with vertex and edge weights (a vertex with higher weight has darker color).]{{\includegraphics[width=10cm]{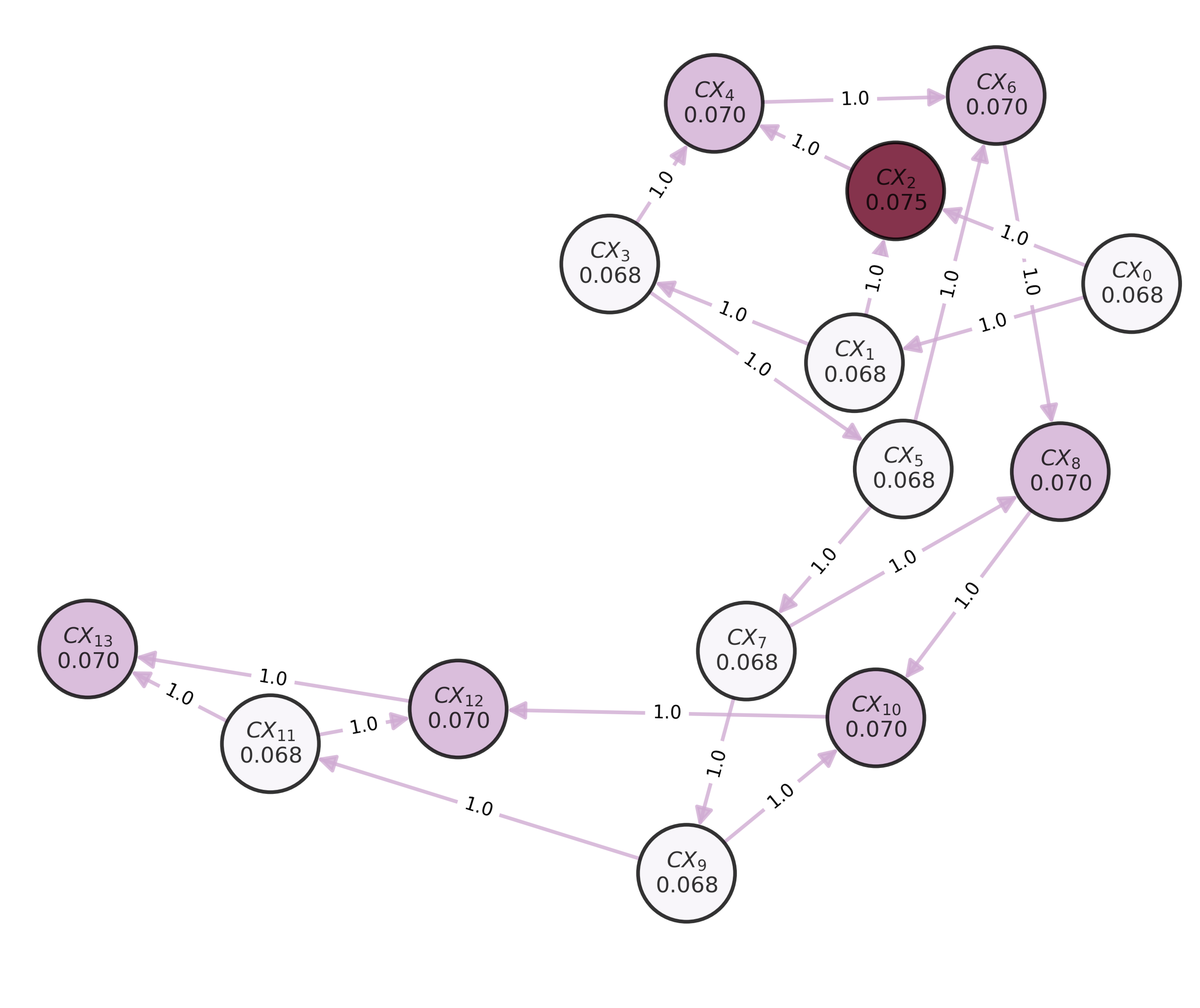} }}%
    \caption{An example circuit and its corresponding doubly-weighted graph.}%
    \label{example circuit}%
\end{figure}

\paragraph{Example}
Let us illustrate with a quantum circuit $C$ having 8 qubits and 14 two-qubit {\it CNOT} ({\it cx}) gates, as shown in Figure  \ref{example circuit}(a). In the corresponding doubly-weighted graph $G_C$,  each vertex represents a {\it CNOT} ({\it CX}) gate and each edge denotes a qubit connecting two {\it CNOT} gate. 
Let us calculate the weight of the vertex $CX13$ which has an edge from $CX11$ and from $CX12$. Hence, while computing the vertex weight, we have to consider:
\begin{enumerate}
    \item the error rate of $CX13$, 
    \item error due to all the single qubit gates applied between $CX11$ to $CX13$,
    \item error for all the single qubit gates operated between $CX12$ to $CX13$
    \item amplitude and phase damping error. 

\end{enumerate}
Thus the weight of $CX13$ is 0.070 as shown in Figure \ref{example circuit}(b). Similarly, we compute the weights for all the vertices as portrayed in Figure \ref{example circuit}(b).

\subsubsection{Error influenced balanced bi-partitioning of circuit}\label{algos}

First, we present the objective function of the optimization problem we intend to solve. 
\paragraph{Objective function} Let us assume that we have a cut $c$, which partitions graph $G_C$ into two sub-graphs $G_{1}$ and $G_{2}$.
Let an indicator variable $y_{v}$ be associated with each vertex $v\in V$ such that,    
\begin{equation}
    y_{v}=
    \begin{dcases}
        0 & if \hspace{2mm}  v\in G_{1} \\
        1 & if \hspace{2mm} v \in G_{2} \\
    \end{dcases}
\end{equation}
If $e_{i,j}$ is the edge connecting the vertices $v_{i}$ and $v_{j}$ having edge weight $w_{i,j}$, then the cut size ($K_c$) for the specific cut $c$, can be written as
\begin{equation}
    K_{c}=\sum_{e_{i,j}\in E}w_{i,j}(y_{v_{i}}-y_{v_{j}})^{2}.
\end{equation}

Let the sum of vertex weights for the partitions $G_{1}$ and $G_{2}$   be $\Omega_{G_{1}}$ and $\Omega_{G_{2}}$ respectively. Hence, the overall cost for a cut $c$ can be written as
\begin{equation}\label{Cost Function}
    Cost_{c}= K_{c} (\frac{1}{\Omega_{G_{1}}}+\frac{1}{\Omega_{G_{2}}}).
\end{equation}

Our aim is to find a cut $c$ for which this $Cost_c$ is minimum. Thus, the cost is given by  Eqn. \ref{Cost Function} is our objective function that has to be minimized. 
In this paper, we apply both classical and quantum approaches to solve this optimization problem to establish the most suitable method. We start with a  genetic algorithm-based proposed classical approach followed by a quantum annealing-based approach.

\paragraph{Proposed classical approach for finding an error-balanced min-cut}
We propose a GA-based approach to minimize our objective function in Eqn. \ref{Cost Function}. Algorithm \ref{algo1} outlines our proposed approach for identifying the minimum cut $c$ in a graph $G_C$ while maintaining a balance with respect to errors.
\begin{algorithm}
\caption{Finding an Error-balanced Min-Cut}\label{algo1}
\begin{algorithmic}
\Require $N$, $initialPartitionVector$ 
\Ensure $MinPartitionVector$, $MinCost$
\Comment{$N =$ number of vertices}
\State $MinVector = initialPartitionVector + 1$
\State $MinCost = CostCalculator(initialPartitionVector)$
\State $CostFlag$ = 0
\While{$CostFlag \leq c_{2}$}
\For{$i \gets 1 \textrm{ to } N$}

$partitionVector[i] = initialPartitionVector[i] \oplus 1$
    
$cost [i] = CostCalculator (partitionVector)$

\If{$cost[i] < MinCost$}
    \State $MinCost = cost[i]$
    \State $MinVector = partitionVector$
    \State $costFlag = 0$
    
    \Else 
    \State $CostFlag = CostFlag + 1$

\EndIf

\EndFor

$initialPartitionVector = crossover(MinVector,initialPartitionVector)$
\EndWhile

\Return{$MinCost$, $MinVector$}

\end{algorithmic}
\end{algorithm}

We initiate the algorithm with a random cut effectively dividing the graph into two sub-graphs. We obtain a partition vector, essentially a string containing the values of $y_{v}$ for all $N$ vertices of $G_C$. Next, the cost of this initial partition is computed by   Algorithm \ref{algo2}, which implements Eqn. \ref{Cost Function}. 
\begin{algorithm}
\caption{Calculating Cost of a Partition}\label{algo2}
\begin{algorithmic}
\Require $cutSize$, $partitionVector$, $vertexWeight$, $N$ 

\Comment{$N =$ number of vertices}
\Ensure Cost
\Function{$CostCalculator$}{$partitionVector$}
\State $WeightP2 \Leftarrow 0$
\For{$i \gets 1 \textrm{ to } N$}
        
          $totalWeight = totalWeight + vertexWeight[i]$
          
          $WeightP2 = WeightP2 + vertexWeight[i] \times partitionVector[i]$

\EndFor
      
    $WeightP1=totalWeight - WeightP2$
    
    $Cost= CutSize(partitionVector) \times (\frac{1}{WeightP1} +\frac{1}{WeightP2})$

\Return{Cost}

\EndFunction

\end{algorithmic}

\end{algorithm}
The procedure, as outlined in Algorithm \ref{algo2}, includes determining the weights of vertices within each partition and employing   Algorithm \ref{algo3} (referred to as the Cut Size Calculator), to compute the size of the partition. Once this cost of the initial partition vector  is computed, it is stored as $MinCost$.

\begin{algorithm}
\caption{Calculating size of min-ut size}\label{algo3}
\begin{algorithmic}
\Require $partitionVector$, $edges$, $edgeWeight$

\Ensure $cutSize$ 

/* $partitionVector$ is '0' or '1' for vertices ($v_{1},v_{2},.. v_{N}$) in Partition1 or Partition2 respectively, edge $e_{k,l}$ have endpoints $v_{k}$ and $v_{l}$, $edgeWeight$ is the weight of each edge */\\

\Function{$CutSize$}{$partitionVector$}
\State $cutSize \Leftarrow 0$

\While{$edges \neq NULL$}        
\Comment{We have considered all the edges}
          
    $cutSize=cutSize + edgeWeight[e_{i,j}] \times (partitionVector[i] - partitionVector[j])^{2}$
    \Comment{Edge $e_{i,j}$ is an edge between vertex $v_{i}$ and $v_{j}$ }
          
\EndWhile    

\Return $cutSize$

\EndFunction

\end{algorithmic}
\end{algorithm}

The subsequent phases of the process iteratively flip each bit in the partition vector and update $MinCost$ if the cost of the new partition vector is lower. This process is repeated until all bits have been flipped. This marks the completion of one iteration or pass through Algorithm 1. Once a pass is completed, Algorithm 1 prepares the initial partition vector for the next iteration from the partition vector with the lowest cost given by $MinCost$ and the initial partition vector, then applies a crossover operation using Algorithm \ref{algo4} to generate the new initial partition vector for the next pass.

\begin{algorithm}[!ht]
\caption{Crossover algorithm}\label{algo4}
\begin{algorithmic}
\Require $Vector1$, $Vector 2$, $N$
\Ensure $FinalVector$
\Function{$crossover$}{$Vector1, Vector2, N, c_{1}$}
\State $FinalVector \Leftarrow 0$

\For{$i \gets 0 \textrm{ to } c_{1}N$}

\State $FinalVector[i] = Vector1[i]$

\EndFor

\For{$j \gets 0 \textrm{ to } N(1-c_{1})$}
    \State $FinalVector[cj+1]=Vector2[j]$
\EndFor

\Return $FinalVector$
\EndFunction
\end{algorithmic}

\end{algorithm}

In summary, the algorithm calculates partition costs, iteratively improves the partition by flipping bits, and uses a crossover operation to create a new initial partition vector for the next iteration, all with the aim of optimizing the partition.
This process is repeated until no further improvement in cost is observed. After reaching this point, the algorithm undergoes a few more iterations (a total of $c_2$ times) before ultimately returning the final $MinCost$ and the corresponding $MinVector$. The overall complexity of our proposed algorithm is $O(n\cdot e)$, where $n$ is the number of vertices and $e$ is the number of edges of the graph.

\paragraph{Quantum Annealing Based Approach}
We incorporate quantum annealing as a quantum approach in {\it FragQC} to solve the proposed optimization problem defined in Eqn. \ref{Cost Function}. In \cite{ushijimamwesigwa2017graph}, the authors have exhibited the graph partitioning using quantum annealing on the D-Wave system. We also use the D-Wave system for our experiments. For minimizing our objective function in Eqn.  \ref{Cost Function} on a D-wave system, we need to describe our objective function in the following Ising objective function
\begin{equation}
\min\left(\sum_i h_i s_i+\sum_{i<j} J_{i j} s_i s_j\right).
\end{equation}
where $s_i \in (+1, -1)$ are subject to local fields $h_i$ and pair wise interactions with coupling strengths $J_{ij}$.

The quadratic unconstrained binary optimization (QUBO) representation is often preferred with its $0/1$-valued variables over the Ising $-1/+1$-valued variables because it is more natural. The QUBO objective function is

\begin{equation}    
\min\left(\sum_i Q_{i i} x_i+\sum_{i<j} Q_{i j} x_i x_j\right).
\end{equation}
where $Q_{ii}$ is analogous to the Ising $h_i$, as are $Q_{ij}$ and $J_{ij}$. The Ising and QUBO models are related through the transformation $s=2x-1$. 

Fortunately, the D-Wave machine allows both the Ising and QUBO forms, hence we describe our objective function as Ising formulation. Our Ising formulation for proposed doubly weighted graph balanced bi-partitioning can be derived from Eqn. \ref{Cost Function},  as follows:
\begin{equation}
\min \left(\sum_{i,j}^{n}{w_{ij}^2}\frac{1-s_i s_j}{2} +  2\sum_{i<j}^{n}{v_i v_j s_i s_j }  +  \sum_{i=1}^{n}{v_i}^{2}  +  \left (\sum_{i=1}^{n}{v_i} - \frac{n}{2}  \right)^{2}\right).
\end{equation}
where $w_{ij}$ is the weight of an edge between vertex $i$ and $j$ and $v_i$ is the weight of a vertex $i$ in $G_C$. The total number of vertices of the graph is $n$. We minimize this objective function through quantum annealing on a D-Wave system to find an optimal cut in a quantum circuit.


\section{Experimental Evaluation}\label{sec4}
We present here the notable results obtained when we fed different benchmark quantum circuits from \cite{Quetschlich2023mqtbench, li2022qasmbench} to our tool {\it FragQC}. We have applied both classical and quantum approaches, i.e., GA-inspired approach (Classical) and quantum annealing-based approach implemented using D-Wave's quantum annealing device. We have compared both the results with {\it h-METIS} which is a very popular and state-of-the-art tool for circuit partitioning.  

 The system configuration on which all the experiments have been performed in Python 3.11.1 with processor AMD EPYC 7B13 (x86\_64) octacore on KVM, CPU 2.5GHz, RAM 62.8Gi Usable, and x86\_64 Ubuntu 22.04.2 LTS Operating System. 
 


\begin{figure}[!ht]
    \centering
    \subfloat[\centering Cost of the cuts selected by {\it h-METIS} vs Proposed QA vs Proposed GA-based approach.]{{\includegraphics[width=6.3cm]{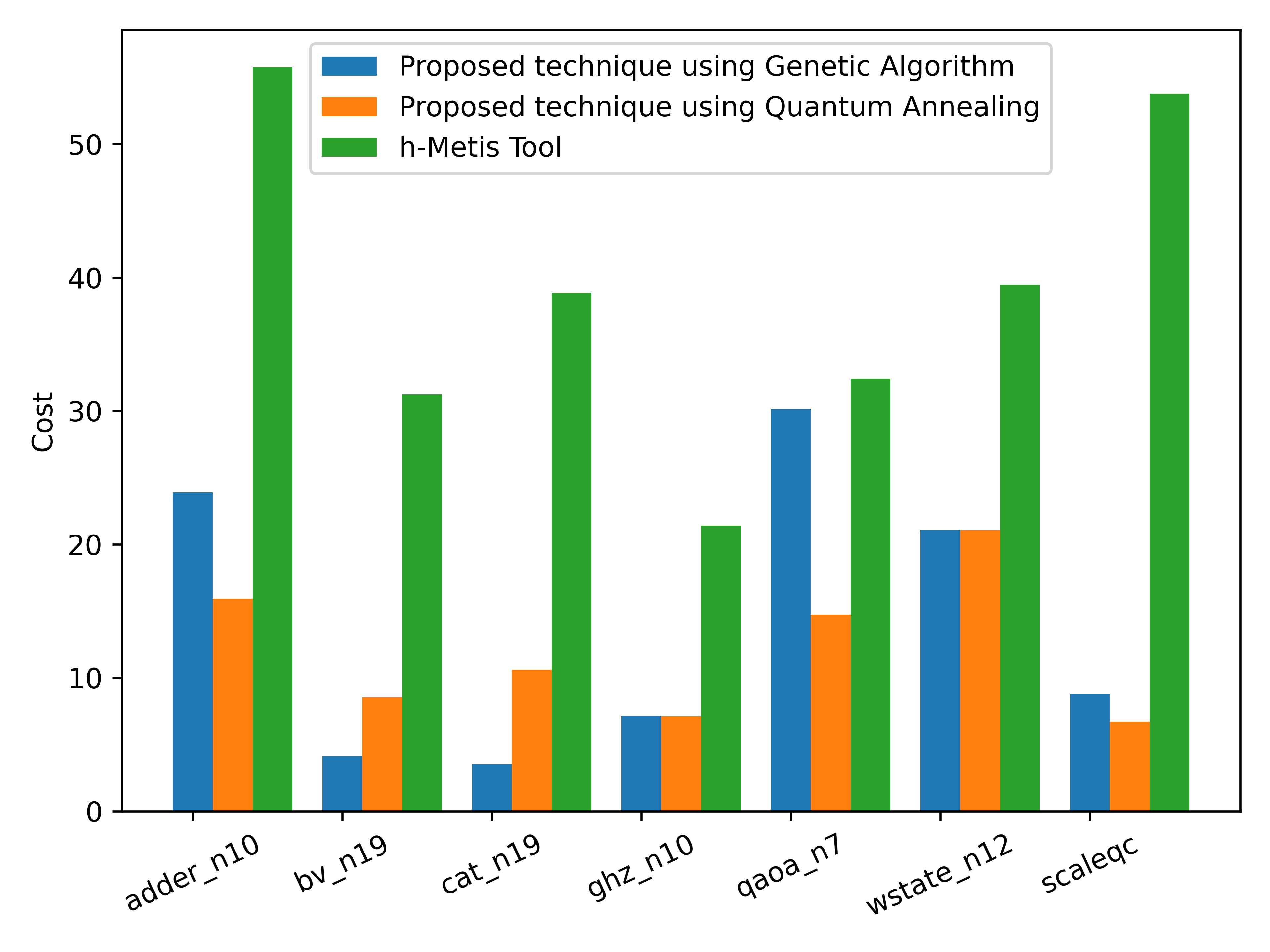} }}%
    \qquad
    \subfloat[\centering  Cut size of the chosen cuts by {\it h-METIS} vs Proposed QA vs Proposed GA-based approach.]{{\includegraphics[width=6.3cm]{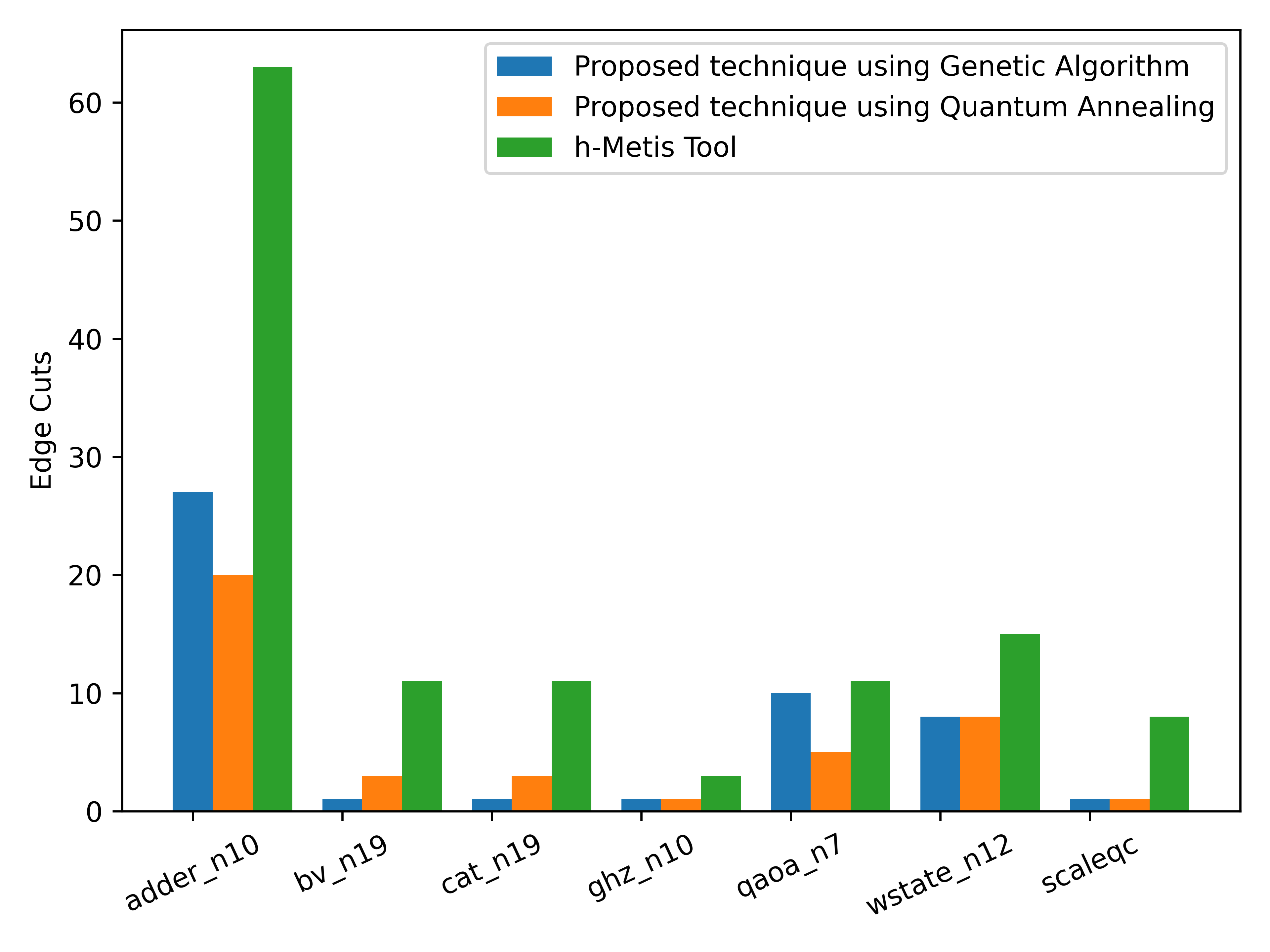} }}%
    \caption{Cost and cut size for the selected cut by {\it h-METIS} (green), Proposed QA (blue), and Proposed GA-based approach (orange) for different benchmark circuits.}%
    \label{cost}%
\end{figure}

Figure \ref{cost}(a) provides the cost of the cut selected by the specified algorithms, and  Figure \ref{cost}(b) the cut sizes. The green bar represents {\it h-METIS}, the orange one is for quantum annealing using the D-Wave systems, and the blue one is for the proposed classical GA-based approach. It is evident from these results that the cut selected by {\it h-METIS} has a much higher cost and cut size, than the other two approaches. However, there is no clear winner between the classical GA-based approach and the quantum annealing-based approach. Hence, for further experiments, we have used  GA-Based and QA-based approaches for our circuits and {\it FragQC} reports the cut that has a lower cost associated with it.      

\begin{figure}[!htb]
    \centering
    \includegraphics[width=10cm]{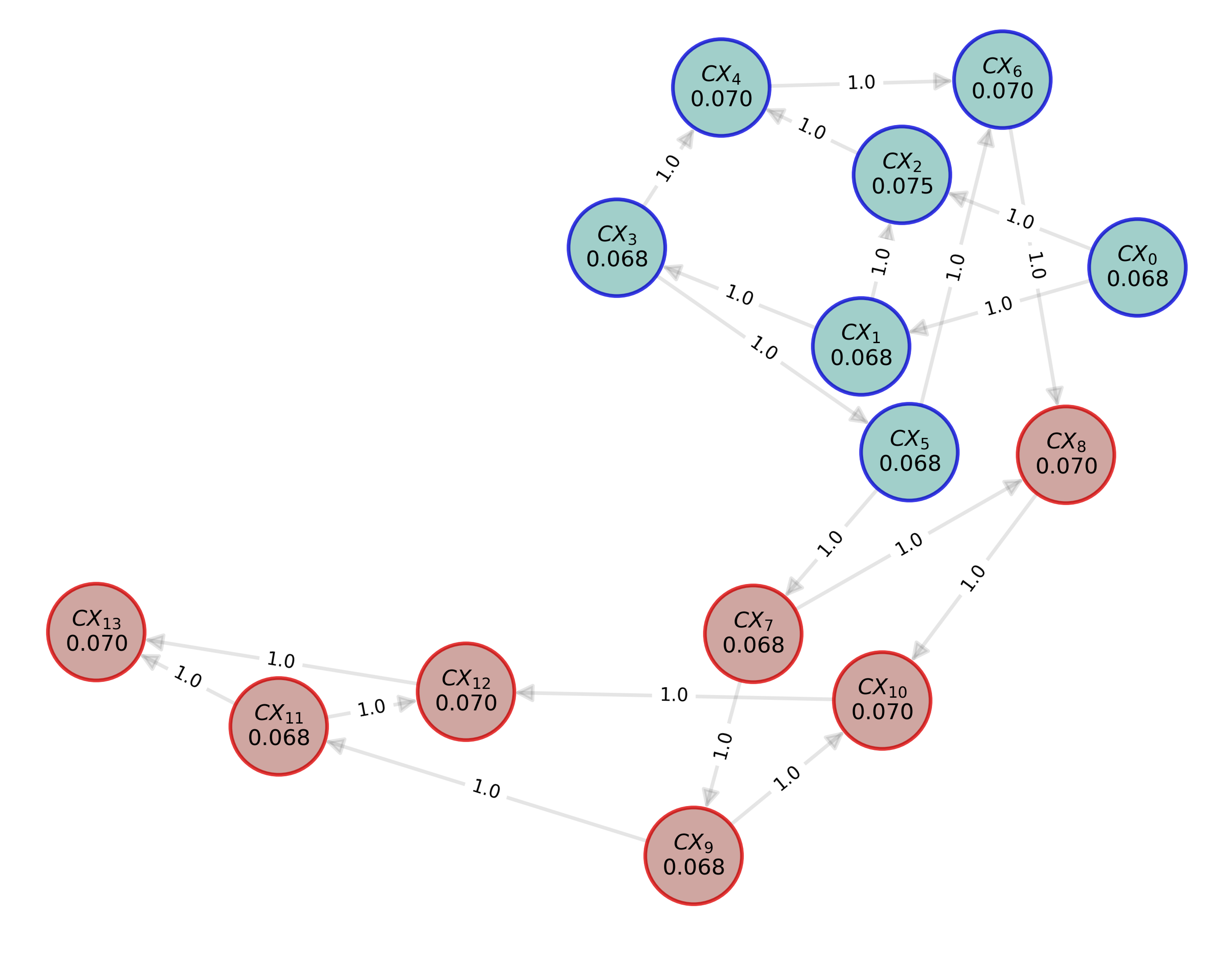}
    \caption{Balance bi-partitioning result for the graph shown in Figure \ref{example circuit}(b) using error balanced min-cut finding algorithm. The vertices of the subgraphs for the two partitions are marked in red and green.}
    \label{partitioning}
\end{figure}

\begin{figure}[!htb]
    \centering
\includegraphics[width=\textwidth]{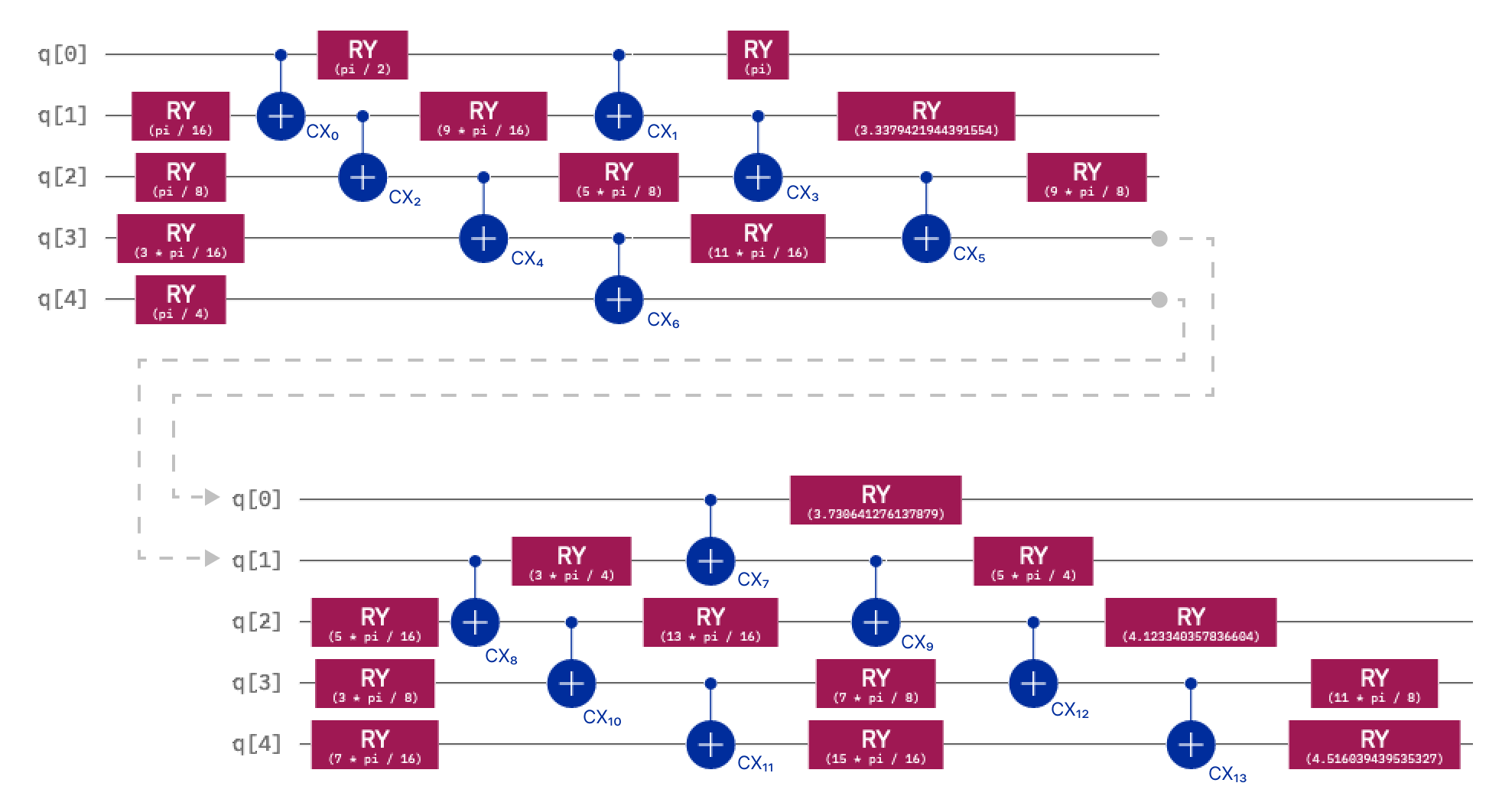}
    \caption{Two sub-circuits corresponding to two sub-graphs of Figure \ref{partitioning} produced by our error balanced Min-cut finding algorithm.}
   \label{subcircuit}
\end{figure}

Let us consider the circuit shown in Figure \ref{example circuit}(a). Its corresponding doubly weighted graph is given in Figure \ref{example circuit}(b). In order to apply our tool {\it FragQC} on this circuit, the error-balanced Min-cut finder reports the partition as shown in Figure \ref{partitioning}. In this case, the blue vertices form one sub-circuit, while the red vertices form the other. For this cut, only 2 edges, each having weight 1, are cut, thus the cut size is 2 and the overall cost for the cut is 8.239. The two sub-circuits are shown in Figure \ref{subcircuit}. These two sub-circuits can then be executed in the quantum hardware and the final outcome can be obtained through classical reconstruction.  

Despite the fact that {\it FragQC} is a hardware-sensitive tool, it can easily be used with gate-based quantum hardware of any technology and size.  We have used our tool {\it FragQC} to cut the quantum circuit whenever the success probability was below a certain user-defined threshold and executed the smaller sub-circuits on quantum hardware. We have calculated the fidelity compared to the ideal simulation. For the purpose of a comparative study, we have leveraged {\it CutQC} \cite{10.1145/3445814.3446758} through the circuit knitting toolbox.  We have also directly executed the circuits on $IBM$\_$Sherbrooke$ without cutting the circuits. The results obtained are shown in the Table \ref{table:fidelity}. For all the circuits, we observed better fidelity for {\it FragQC} over {\it CutQC} and direct execution without cutting. On average, we got 14.83\% better fidelity compared to direct execution without cutting the circuit and 8.45\% fidelity gain over {\it CutQC}.    

\begin{table}[!ht]
\centering
\caption{Fidelity for benchmark circuits.}
\begin{tabular}{|c|c|c|c|c|c|}
\hline
\multirow{3}{*}{Benchmark circuit} & \multirow{3}{*}{Width} & \multirow{3}{*}{Depth}  & \multicolumn{3}{c|}{Fidelity} \\
\cline{4-6}
& & & \multicolumn{1}{c|}{Without cut} & \multicolumn{1}{c|}{\it CutQC} & \multicolumn{1}{c|}{{\it FragQC}} \\
\cline{4-6}
\hline
Efficient SU & 8 &  12 & 0.849  & 0.864  & 0.879
\\\hline
ghz\_n10 &  10 &  11 &0.738  & 0.735  & 0.799 
\\\hline 
Adder n\_10 &  10 &  108 & 0.735  & 0.734  & 0.738 
\\\hline
bv\_n19 &  20 &  22 & 0.497  & 0.534  & 0.598 
\\\hline 
cat\_n24 & 24 &  25 & 0.394 & 0.491  & 0.558 
\\\hline

\end{tabular}

\label{table:fidelity}
\end{table}

Since quantum circuit fragmentation and knitting involve several measurement operations, we wanted to eradicate errors that occurred due to noisy measurement operations. For this purpose, we have used IBM's measurement error mitigation process to reduce the readout errors from the probability distribution output generated by each sub-circuit after hardware execution. Table \ref{table:fidelitywithEM} displays the fidelity of the benchmark circuits when executed on {\it FragQC} along with measurement error mitigation. We have also integrated measurement error mitigators with {\it CutQC} and compared their corresponding fidelity. The fidelity obtained using {\it FragQC} is approximately 8.99\% better than the {\it CutQC} with error mitigation.

\begin{table}[!ht]
\centering
\caption{Fidelity for benchmark circuits with Measurement Error Mitigation.}
\begin{tabular}{|c|c|c|c|c|}
\hline
\multirow{3}{*}{Benchmark circuit} & \multirow{3}{*}{Width} & \multirow{3}{*}{Depth}  & \multicolumn{2}{c|}{Fidelity} \\
\cline{4-5}
& & & \multicolumn{1}{c|}{\it CutQC} & \multicolumn{1}{c|}{{\it FragQC}}  \\
\cline{4-5}
\hline
Efficient SU & 8 &  12   & 0.874 & 0.898
\\\hline
ghz\_n10 & 10 &  11   & 0.951 & 0.986
\\\hline 
Adder n\_10 &  10 &  108   & 0.813  & 0.957
\\\hline
bv\_n19 & 20 &  22   & 0.861 & 0.963
\\\hline 
cat\_n24 & 24 &  25   & 0.8997  & 0.989
\\\hline

\end{tabular}

\label{table:fidelitywithEM}
\end{table}

\begin{figure}[!ht]
\centering
\includegraphics[scale=0.03]{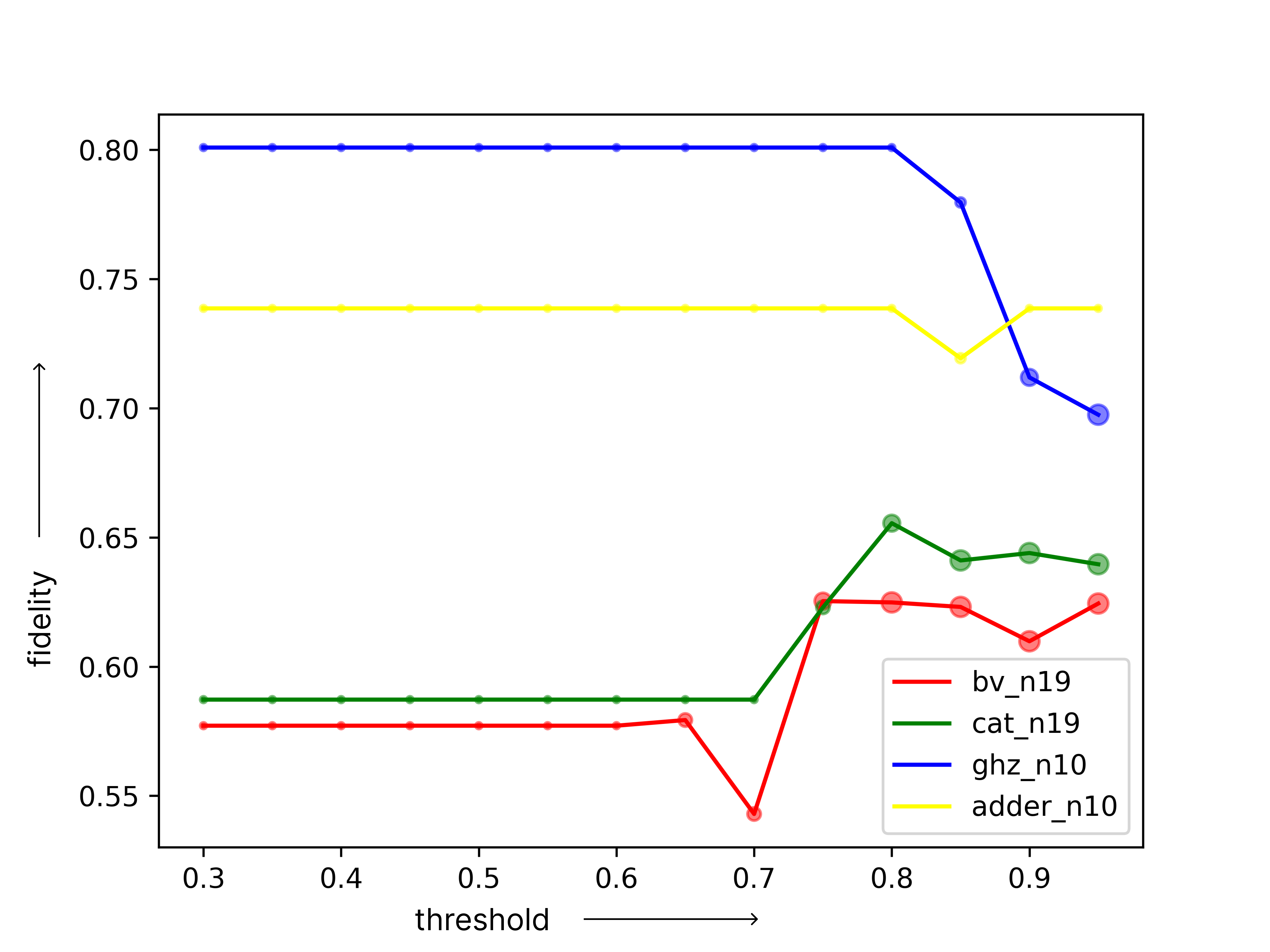}
\caption{Fidelity obtained by using {\it FragQC} on four benchmark circuits,  with varying thresholds of success probability.}
\label{threshold}
\end{figure}

As mentioned earlier, the determination of the threshold for {\it FragQC} for benchmark circuits is specified by the user. Albeit, we try to show through numerical analysis that it would not guarantee that the fidelity for benchmark circuits would always increase if the threshold level is set very high. We have taken four example benchmark circuits to analyze the fidelity of the circuit using {\it FragQC}, while we vary the success probability threshold. In Figure \ref{threshold}, we portray the numerical results from which we can conclude that increasing the success probability threshold would not ensure an increase in fidelity.



\section{Conclusion}\label{sec5}
In this paper, we have proposed a method for balanced bi-partitioning of a doubly weighted graph for quantum circuit fragmentation by considering both the hardware noise and the cut size. We have also exhibited through our proposed tool, {\it FragQC}, that the fidelity of the benchmark circuits has significantly improved with a GA-based method or a quantum annealing method within it, for solving the problem in comparison with the existing circuit fragmentation methods. Therefore, \textit{FragQC} provides a hybrid quantum computing strategy. It is also a robust and scalable method, as this method can be implemented on any gate-based hardware. Since there may be multiple numbers of sub-circuits, these may be implemented and run on different hardware in parallel \cite{bhoumik2023distributed, 9334411, Chatterjee_2022} to minimize the run-time complexity of our tool. This may be explored in the future along with the trade-off in the time for reconstruction of the results of the sub-circuits.

\section*{Declaration of competing interest}
The authors declare that there is no conflict of interest. All authors have contributed equally in this manuscript.

\small
\bibliographystyle{elsarticle-num} 
\bibliography{sn-bibliography}





\end{document}